\documentclass{article}

\normalsize
\advance\textheight by \topskip
\twocolumn

\usepackage{times}
\usepackage{graphicx} 
\def\lsi{\raise0.3ex\hbox{$<$\kern-0.75em\raise-1.1ex\hbox{$\sim$}}}
\def\gsi{\raise0.3ex\hbox{$>$\kern-0.75em\raise-1.1ex\hbox{$\sim$}}}
\newcommand{\lsim}{\mathop{\lsi}}

\begin{document}

\title{Ultrahigh energy cosmic rays as a Grand Unification signal} 
\author{Z. Fodor and S.D. Katz\\
Institute for Theoretical Physics, E\"otv\"os University, \\P\'azm\'any
1, H-1117 Budapest, Hungary}


\maketitle

\begin{abstract}
We analyze the spectrum of the ultrahigh energy (above $\approx 10^{9}$~GeV)
cosmic rays. With a maximum likelihood  analysis we show that the
observed spectrum is consistent with the decay of
extragalactic GUT scale particles. The
predicted mass for these superheavy particles is 
$m_X=10^b$~GeV, where $b=14.6_{-1.7}^{+1.6}$.
\end{abstract}

\section{Introduction}

The interaction of protons with photons of the cosmic microwave 
background radiation (CMBR) predicts a sharp drop in the cosmic ray flux above
the Greisen-Zatsepin-Kuzmin (GZK) cutoff around $5\cdot 10^{19}$~eV 
(Greisen, 1966; Zatsepin and Kuzmin, 1966). The 
available data shows no such drop. About 20 events above $10^{20}$~eV
were observed by experiments such as AGASA 
(Takeda et al., 1998), Fly's Eye (Bird et al., 1993), 
Haverah Park (Lawrence et al., 1991),
Yakutsk (Efimov et al., 1991) and HiRes (Kieda et al., 1999). 
Future experiments, particularly Pierre Auger (Boratav, 1996; 
Guerard, 1999; Bertou et al. 2000), will 
have a much higher statistics.

Usually it is assumed that 
at these energies the galactic and extragalactic (EG) magnetic 
fields do not affect the orbit of the cosmic rays, thus they 
should point back to their origin within a few degrees. 
Though there are clustered events (Hayashida et al., 1996; 
Uchihori et al., 2000)
the overall distribution is practically isotropic
(Dubovski and Tinyakov, 1998; Berezinsky and Mikhailov, 1999;
Medina Tanco and Watson, 1999),
which usually ought to be interpreted as a signature for EG 
origin. 
Since above
the GZK energy the attenuation length of particles is a few tens
of megaparsecs (Yoshida and Teshima, 1993 ; Aharonian and Cronin, 1994;
Protheroe and  Johnson, 1996; Bhattacharjee and  Sigl, 2000; 
Achterberg et al., 1999; T. Stanev et al., 2000)
if an ultrahigh energy cosmic ray (UHECR) is
observed on Earth it must be produced in our vicinity. 
Sources of EG origin should 
result in a GZK cutoff, which is in disagreement with experiments.  
It is generally believed that there is no 
conventional astrophysical explanation for the observed UHECR spectrum.

An interesting idea is that 
superheavy particles (SP) as dark matter could be the source of UHECRs. 
In (Kuzmin and Rubakov, 1998) EG SPs were studied.
A crucial observation was made (Berezinsky et al., 1997) about the
decay of SPs concentrated in the 
halo of our galaxy. They used  the modified leading logarithmic
approximation (MLLA) (Azimov et al., 1985; Fong and Webber 1991) 
for ordinary QCD and for supersymmetric QCD
(Berezinsky and Kachelrie{\ss}, 1998). 
A good agreement of the EG spectrum with observations was noticed in 
(Berezinsky et al., 1998).
Supersymmetric QCD is treated as the strong regime of the
minimal supersymmetric standard model (MSSM).
To describe the decay spectrum more 
accurately HERWIG Monte-Carlo (Marchesini et al., 1992)
was used in QCD (Birkel and Sarkar, 1998) and discussed
in supersymmetric QCD (Rubin 1999, Sarkar 2000), resulting in 
$m_X \approx 10^{12}$ GeV and $\approx 10^{13}$ GeV for the SP mass in  
SM and in MSSM, respectively. 

SPs are very 
efficiently produced by the various mechanisms at post inflatory 
epochs (for a review see Berezinsky 2000). 
Note, that any analysis of SP decay 
covers a much broader class of possible sources. 
Several non-conventional UHECR sources
produce the same UHECR spectra as decaying SPs. 

Here we study the scenario that the UHECRs are coming
from decaying SPs and we determine the mass of this $X$ 
particle $m_X$ by a detailed analysis of the observed UHECR spectrum. 
We discuss both possibilities that the UHECR protons are 
produced in the halo of our galaxy and that they are of EG
origin and their propagation is affected by CMBR. Here we do not investigate
how can they be of halo or EG origin, 
we just analyze their effect on the observed
spectrum instead. We assume that the SP decays into 
two quarks (other decay modes 
would increase $m_X$ in our conclusion). 
After hadronization these quarks yield protons. The result is characterized 
by the fragmentation function (FF) $D(x,Q^2)$ 
which gives the number of produced 
protons with momentum fraction $x$ at energy scale $Q$.
For the proton's FF present accelerator
energies can be used 
(Binnenwies et al., 1995;  Kniehl et al., 2000). We evolve
the FFs in ordinary (Gribov and Lipatov 1972; 
Lipatov 1975; Altarelli and Parisi, 1977;
Dokshitzer, 1977) and 
in supersymmetric (Jones and Llewellyn Smith 1983) QCD to the energies of the 
SPs. This result can be combined with the
prediction of the MLLA technique 
, which gives 
the initial spectrum of UHECRs at the energy $m_X$. 
Altogether we study four different models:
halo-SM, halo-MSSM, EG-SM and EG-MSSM. 

\section{Decay and fragmentation of heavy particles}

The UHECRs are most likely
to be dominated by protons (Dawson et al., 1998); in 
our analysis we use them exclusively.

The FF of the proton can be determined from 
present experiments (Binnenwies et al., 1995;  Kniehl et al., 2000). 
The FFs at $Q_0$ energy scale are $D_i(x,Q_0^2)$,
where $i$ represents the different partons (quark, squark or gluon, gluino).
The FFs can not be determined
in perturbative QCD; however, their evolution in $Q^2$ 
is governed by the DGLAP equations 
(Gribov and Lipatov 1972; Lipatov 1975; Altarelli and Parisi, 1977;
Dokshitzer, 1977):
\begin{eqnarray} 
{\partial D_i(x,Q^2) \over \partial \ln Q^2}= 
\nonumber \\ 
\frac{\alpha_s(Q^2)}{2\pi}
\sum_j \int_x^1 {dz \over z} P_{ji}(z,\alpha_s(Q^2))D_j(\frac{x}{z},Q^2),
\end{eqnarray} 
where 
$P_{ji}(z)$ is the splitting function.
We solve the DGLAP equations numerically with the 
conventional QCD (SM case) splitting functions and with the supersymmetric 
(MSSM ca\-se) ones (Jones and Llewellyn Smith 1983). 
For the top and the MSSM partons
we used the FFs of ref. (Rubin 1999). While solving the DGLAP equations each 
parton is included at its own threshold energy.

\begin{table}[t]
\begin{center}\begin{tabular}{l|l|l|l|l}
flavor&Q(GeV)&$N$&$\alpha$&$\beta$\\
\hline
$u=2d$ &1.41&0.402&-0.860&2.80\\
\hline
$s$    &1.41&4.08&-0.0974&4.99\\
\hline
$c$    &2.9&0.111&-1.54&2.21\\
\hline
$b$    &9.46&40.1&0.742&12.4\\
\hline
$t$    &350&1.11&-2.05&11.4\\
\hline
$g$    &1.41&0.740&-0.770&7.69\\
\hline
$\tilde{q}_i, \tilde{g}$&1000&0.82&-2.15&10.8\\
\end{tabular}
\vspace{0.3cm}
\caption{\label{frag_tab}
{The fragmentation functions of the different partons using the 
parametrization $D(x)=Nx^\alpha (1-x)^\beta$ at different energy scales
(second column).
}}
\end{center}\end{table}

\begin{figure}[t]
\vspace*{2.0mm}
\includegraphics[width=8.3cm]{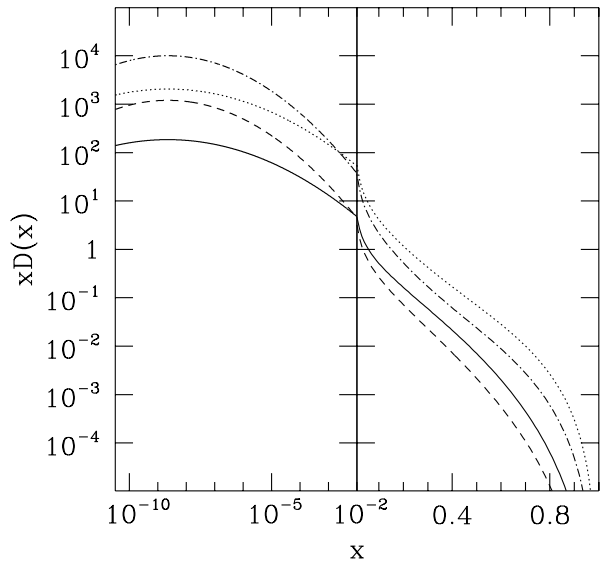}
\caption{\label{fragmentation}
{The FFs averaged over the quark flavors
at $Q=10^{16}$ GeV for proton/pion in SM (solid/dotted line)  and in MSSM 
(dashed/dashed-dotted line)
in the relevant $x$ region. To show both the small and large $x$
behavior we change from logarithmic scale to linear at $x=0.01$.
}}
\end{figure}

At small values of $x$, multiple soft gluon emission can be 
described by the MLLA.
It describes the observed hadroproduction quite accurately
in the small $x$ region (see eg. Abreu et al., 1999; Abbiendi et al., 2000).
For large values of $x$ the MLLA should not be
used. We smoothly connect the solution for the FF
obtained by the DGLAP equations 
and the MLLA result at a given $x_c$ value.
Our final result on $m_X$
is rather insensitive to the choice of $x_c$, the uncertainty is included
in our error estimate. We also determined the FF of the pion.
Fig. \ref{fragmentation} shows the FF for the proton and 
pion at $Q=10^{16}$~GeV in  SM and MSSM.

\section{Comparison of the predicted and the \\observed spectra}

UHECR protons produced in the halo of our galaxy can propagate
unaffected and the production spectrum should be
compared with the observations. 

Particles of EG origin
and energies above $\approx 5\cdot 10^{19}$ eV loose a large fraction 
of their energies due to interactions with CMBR 
(Greisen, 1966; Zatsepin and Kuzmin, 1966). This effect can be quantitatively described by the
function $P(r,E,E_c)$, the probability
that a proton created at a distance $r$ with energy $E$ arrives
at Earth above the threshold energy $E_c$ (Bahcall and Waxman 2000).
This function has been calculated for a wide range
of parameters in (Fodor and Katz, 2001a), 
which we use in the present calculation.
The original UHECR spectrum 
is changed at least by two different ways: (a) there should be a 
steepening due to the GZK effect; (b) particles loosing their
energy are accumulated just before the cutoff and produce a bump. 
We study the observed spectrum by
assuming a uniform source distribution for UHECRs.

\begin{figure}[t]
\vspace*{2.0mm}
\includegraphics[width=8.3cm]{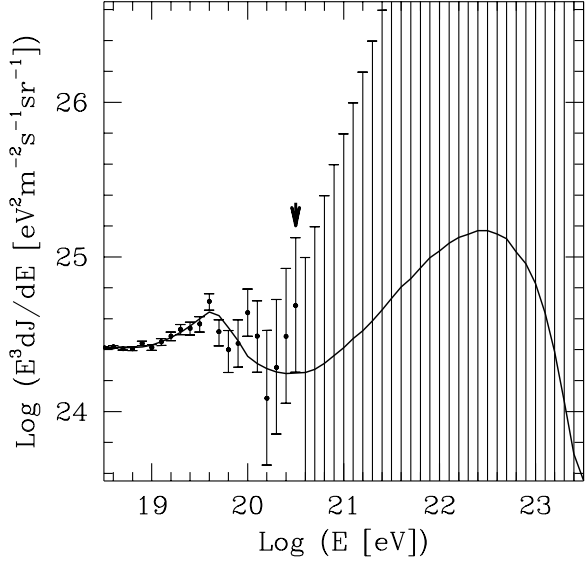}
\caption{\label{spect}
{The available UHECR data with their error bars
and the best fit from a decaying SP.
Note that there are no events above $3 \times 10^{20}$~eV 
(shown by an arrow). 
Nevertheless the experiments are sensitive even in this region. Zero event
does not mean zero flux, but a well defined upper bound for the flux 
(given by the Poisson distribution).
Therefore the experimental value of the
integrated flux is in the ''hatched'' region with 68\% confidence level. 
(''hatching'' is a set of individual 
error bars; though most of them are too large to be depicted in full) 
Clearly, the error bars are large enough to be consistent with the SP decay.
}}
\end{figure}

Our analysis includes the published and the unpublished
(from the www pages of the experiments) UHECR data 
of AGASA, Fly's Eye, Haverah Park and HiRes. Since the decay of SPs results 
in  a non-negligible flux for lower energies $\log (E_{min}/\mbox{eV})=18.5$ 
is used as a lower end for the UHECR spectrum. Our results are
insensitive to the definition of the upper end (the flux is
extremely small there) for which we choose $\log (E_{max}/\mbox{eV})=26$.
As it is usual we divided each logarithmic unit into ten bins. 
Using a Monte-Carlo method
we included this uncertainty in the final error estimates.
The predicted number of events in a bin is given by
\begin{equation}\label{flux}
N(i)=\int_{E_i}^{E^{i+1}}\left[A \cdot E^{-3.16}+B\cdot j(E,m_X)\right],
\end{equation}
where $E_i$ is the lower bound of the i$^{th}$ energy bin. The first 
term describes the data below $10^{19}$~eV according to
(Takeda et al., 1998), where the SP decay gives negligible contribution.
The second one corresponds to the spectrum of the decaying
SPs.
A and B are normalization factors.

The expectation value for the number of events in a bin is given
by eqn. (\ref{flux}) and it is Poisson distributed. To
determine the most probable $m_X$ value we used the maximum-likelihood
method by minimalizing the $\chi^2(A,B,m_X)$ for
Poisson distributed data (Groom et al., 2000)
\begin{equation} \label{chi}
\chi^2=\sum_{i=18.5}^{26.0}
2\left[ N(i)-N_o(i)+N_o(i)\ln\left( N_o(i)/N(i)\right) \right],
\end{equation}
where $N_o(i)$ is the total number of observed events in the i$^{th}$ 
bin. In our fitting procedure we have three parameters: $A,B$ and $m_X$.
Fig. \ref{spect} shows the measured UHECR spectrum and the best fit in the
EG-MSSM scenario.
The first bump of the fit represents particles produced at
high energies and accumulated just above the GZK cutoff due to their energy
losses. The bump at higher energy is a remnant of $m_X$. In the halo
models there is no GZK bump, so the relatively large $x$ part of the FF moves
to the bump around $5\times 10^{19}$~GeV resulting in a much smaller $m_X$ than
in the EG case. 
The experimental data is far more accurately described by the 
GZK effect (dominant feature of the EG fit) than by the FF itself (dominant for 
halo scenarios).

\section{Results}

\begin{figure}
\vspace*{2.0mm}
\includegraphics[width=8.3cm]{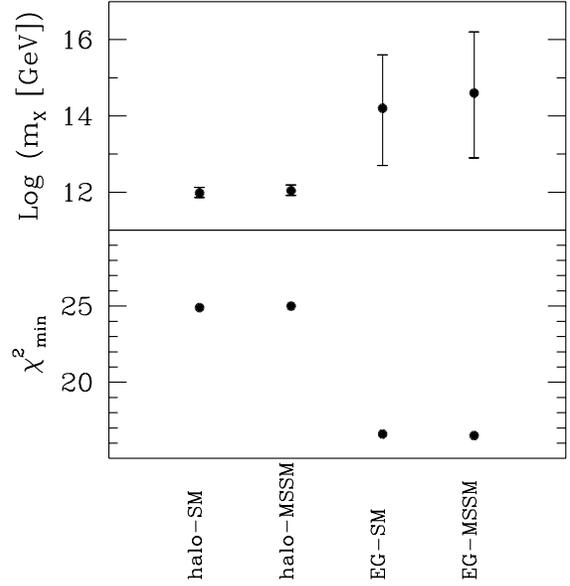}
\caption{\label{result}
{The most probable values for the mass of the decaying
ultra heavy dark matter with their error bars and the
total $\chi^2$ values. Note that 21 bins contain nonzero number of events
and eqn.(\ref{flux}) has 3 free parameters.
}}
\end{figure}

To determine the most probable value for the mass of the 
SP we studied 4 scenarios. Fig. \ref{result} contains
the $\chi^2_{min}$ values and the most
probable masses with their errors for these scenarios.

The UHECR data favors the EG-MSSM scenario. The
goodnesses of the fits for the halo models are far worse. 
The SM and MSSM cases do not differ significantly. 
The most important message is that the masses of the best fits 
(EG cases)
are compatible within the
error bars with the MSSM gauge coupling unification GUT scale 
(Amaldi at al., 1991).

The SP decay will also produce a huge number of pions which will decay into 
photons. Our spectrum contains 94\% of pions and 6\% of protons.
This $\pi /p$ ratio is in agreement with
(Bhattacharjee and  Sigl, 2000; Bhattacharjee and  Sigl, 2000) which showed that
for different classes of models $m_X \lsim 10^{16}$~GeV , which is the upper boundary
of our confidence intervals, the generated
gamma spectrum is still consistent with the observational constraints. 
We performed the whole analysis including the pion produced $\gamma$-s in eqn.
(\ref{chi}). The results agree with our results of Fig. \ref{result} within 
errorbars.

In the near future the UHECR statistics will probably be increased by an
order of magnitude (Boratav, 1996; Guerard, 1999; Bertou et al. 2000). 
Performing our analysis for such a
statistics the uncertainty of $m_X$ was found to
be reduced by two orders of magnitude.

Since the decay time should be at least
the age of the universe it might happen that such SPs
overclose the universe. Due to the large mass of the SPs a single decay 
results in a large number of UHECRs, thus a relatively small 
number of SPs can describe the observations. We checked that in 
all of the four scenarios the minimum density required for the 
best-fit spectrum is more 
than ten orders of magnitude smaller than the critical one.

More details of the present analysis can be found in 
(Fodor and Katz, 2001b). Note, that a similar study
based on the Z-burst scenario (Fargion et al., 1999; Weiler,
1999) can be carried out which gives the mass of the heaviest neutrino
(Fodor et al., 2001).

We thank B.A. Kniehl for providing us with the proton's FF prior
to its publication and F. Csikor for useful comments.
This work was partially supported by Hung. Sci. Foundation
grants No. OTKA-T29803/\-T22929-FKP-0128/\-1997/\-OMMU708/\-IKTA/\-NIIF.

\vspace*{0.5cm}\noindent
{\Large {\bf References}}{\small
\begin{description}
\item
Abbiendi,G. et al., hep-ex/0002012.
\item
Abreu,P. et al., Phys. Lett. {\bf B459}, 397 (1999).
\item
Achterberg,A. et al., astro-ph/9907060.
\item
Aharonian,F.A. and Cronin,J.W., Phys. Rev. {\bf D50}, 1892 (1994).
\item
Altarelli,G. and Parisi,G., Nucl. Phys. {\bf B126}, 298 (1977).
\item
Amaldi,U., de Boer,W., and Furstenau,H.,  Phys. Lett. {\bf B260}, 447 (1991).
\item
Azimovi,Ya.I. et al., Phys. Lett. 
{\bf B165}, 147 (1985); Z. Phys. {\bf C27}, 65 (1985);
$ibid$ {\bf C31}, 213 (1986).
\item
Bahcall,J.N., and Waxman,E., Astrophys.J. {\bf 541}, 707 (2000).
\item
Berezinsky,V., astro-ph/0001163.  
\item
Berezinsky,V., Blasi,P., and Vilenkin,A., Phys. Rev. {\bf D58},
103515 (1998).
\item
Berezinsky,V. and Kachelrie{\ss},M., Phys.Lett. {\bf B434}, 61 (1998).
\item
Berezinsky,V., Kachelrie{\ss},M.,   and Vilenkin,A., 
Phys. Rev. Lett. {\bf 79}, 4302 (1997).
\item
Berezinsky,V. and Mikhailov,A.,A., Phys. Lett. {\bf B449}, 61 (1999).
\item
Bertou,X., Boratav,M., and Letessier-Selvon,A., astro-ph/0001516.
\item
Bhattacharjee,P., and Sigl,G., Phys. Rep. {\bf 327}, 109 (2000).
\item
Binnenwies,J., Kniehl,B.A., and Kramer,G., Phys. Rev. {\bf D52}, 4947 (1995).
\item
Bird,D.J. et al., Phys. Rev. Lett. {\bf 71}, 3401 (1993);
Astrophys J. {\bf 424}, 491 (1994); ibid {\bf 441}, 144 (1995).
\item
Birkel,M. and Sarkar,S., Astropart. Phys. {\bf 9}, 297 (1998).
\item
Boratav,M.,  Nucl. Phys. Proc. {\bf 48}, 488 (1996).
\item
Dawson,B.R., Meyhandan,R., and Simpson,K.M., 
Astropart. Phys. {\bf 9}, 331 (1998).
\item
Dokshitzer,Yu.L., Sov. Phys. JETP {\bf 46}, 641 (1977).
\item
Dubovski,S.L. and Tinyakov,P.G., JETP Lett. {\bf 68}, 107 (1998).
\item
Efimov,N.N. et al., "Proc. Astrophysical 
Aspects of the Most Energetic Cosmic Rays", p. 20, eds. 
M.~Nagano and F.~Takahara, World Sci., Singapore, 1991.
\item
Fargion,D., Mele,B., and Salis,A., Astrophys. J. 517 (1999) 725.
\item
Fodor,Z. and  Katz,S.D., Phys. Rev. {\bf D63}, 023002 (2001).
\item
Fodor,Z. and  Katz,S.D., Phys. Rev. Lett. {\bf 86}, 3224 (2001).
\item
Fodor,Z., Katz.S.D., and Ringwald,A., hep-ph/0105064.
\item
Fong,C.P. and Webber,B.R., Nucl. Phys.  {\bf B355}, 54 (1991).
\item
Greisen,K., Phys.Rev.Lett. {\bf 16}, 748 (1966). 
\item
Gribov,V.N. and Lipatov,L.N., Sov. J. Nucl. Phys. {\bf 15}, 438 (1972). 
\item
Groom,D.E. et al., Eur. Phys. J. {\bf C15}, 1 (2000).
\item
Guerard,C.K., ibid {\bf 75A}, 380 (1999).
\item
Hayashida,N. et al., Phys. Rev. Lett. {\bf 77}, 1000 (1996).
\item
Kieda,D. et al., to appear in Proc. of the 26th ICRC, Salt Lake,
1999; www.physics.utah.edu/Resrch.html
\item
Kniehl,B.A., Kramer,G., and Potter,B., Phys. Rev. Lett. {\bf 85}, 5288 (2000);
Nucl. Phys. {\bf B582}, 514 (2000);
\item
Jones,S.K. and Llewellyn Smith,C.H., Nucl. Phys. {\bf B217}, 145 (1983).
\item
Kuzmin,V.A. and Rubakov,V.A., Phys. Atom. Nucl. {\bf 61}, 1028 (1998).
\item
Lawrence,M.A., Reid,R.J.O., and Watson,A.A.,
J. Phys. {\bf G17}, 773 (1991).
\item
Lipatov,L.N., $ibid$ {\bf 20}, 94 (1975). 
\item
Marchesini,G. et al., Comp. Phys. Comm. {\bf 67}, 465 (1992). 
\item
Medina Tanco,C.A. and Watson,A.A., Astrop. Phys. {\bf 12}, 25 (1999).
\item
Protheroe,R.J., Johnson,P., Astropart. Phys. {\bf 4}, 253 (1996).
\item
Rubin,N., www.stanford.edu/$^\sim$nrubin/Thesis.ps
\item
Sarkar,S., hep-ph/0005256.
\item
Sigl,G. at al., Phys. Rev. {\bf D59}, 043504 (1999).
\item
Stanev,T. et al., astro-ph/0003484.
\item
Takeda,M. et al., Phys. Rev. Lett. {\bf 81}, 1163 (1998); astro-ph/\-9902239.
\item
Uchihori,Y. et al., Astropart. Phys. {\bf 13}, 151 (2000).
\item
Yoshida,S., Teshima,M., Prog. Theor. Phys. {\bf 89}, 833 (1993).
\item
Weiler,T.J., Astropart. Phys. {\bf 11} (1999) 303; Astropart. Phys. {\bf 12}
(2000) 379 (Erratum).
\item
Zatsepin,G.T. and Kuzmin,V.A., Pisma Zh.Exp.Teor.Fiz. {\bf 4}, 114 (1966).
\end{description}}
\end{document}